# Effect of incoherent electron-hole pairs on high harmonic generation in an atomically thin semiconductor

Kohei Nagai[1], Kento Uchida[1], Satoshi Kusaba[1], Takahiko Endo[2], Yasumitsu Miyata[2], and Koichiro Tanaka[1,3]

[1]Department of Physics, Graduate School of Science, Kyoto University, Sakyo-ku, Kyoto 606-8502, Japan

[2]Department of Physics, Tokyo Metropolitan University, Hachioji, Tokyo 192-0397, Japan

[3]Institute for Integrated Cell-Material Sciences, Kyoto University, Sakyo-ku, Kyoto, 606-8501, Japan

**Abstract**

High harmonic generation (HHG) in solids reflects the underlying nonperturbative nonlinear dynamics of electrons in a strong light field and is a powerful tool for ultrafast spectroscopy of electronic structures. Photo-carrier doping allows us to understand the carrier dynamics and the correlations between the carriers in the HHG process. Here, we study the effect of incoherent electron-hole pairs on HHG in an atomically thin semiconductor. The experimentally observed response to photo-carrier doping is successfully reproduced in numerical simulations incorporating the photo-excited carrier distribution, excitonic Coulomb interaction and electron-electron scattering effects. The simulation results reveal that the presence of photo-carriers enhances the intraband current that contributes to high harmonics below the absorption edge. We also clarify that the excitation-induced dephasing process rather than the phase-space filling effect is the dominant mechanism reducing the higher order harmonics above the absorption edge. Our work provides a deeper understanding of high harmonic spectroscopy and the optimum conditions for generating extreme ultraviolet light from solids.

**I. Introduction**

Non-perturbative light-matter interactions of electrons in solids reveal a variety of coherent dynamics that cannot be understood within the framework of perturbation theory [1,2]. High harmonic generation (HHG), which converts low-energy photons to visible or ultraviolet photons, has opened up a new route to examine the non-perturbative dynamics of electrons [3–10]. HHG has been observed by irradiating semiconductors or insulators with strong mid-infrared (MIR) driving fields. Since HHG reflects the underlying non-perturbative dynamics in solids, it can be used to probe various material properties, such as the band structure, Berry curvature, dipole moment structure, atomic bonding, and valence electron density [6,10–14].

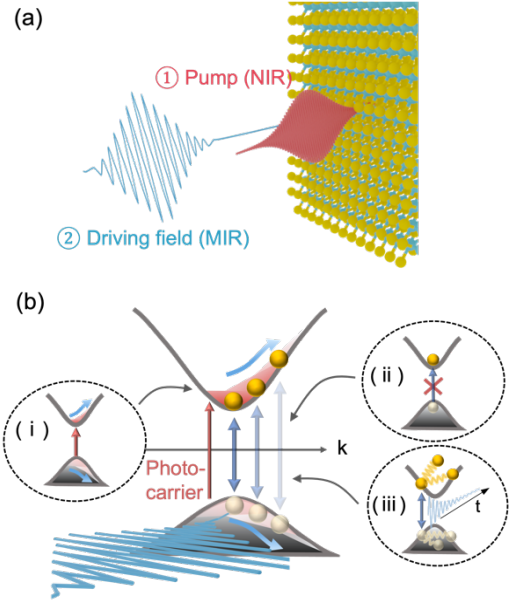

Fig. 1 Effect of photo-carrier doping on high harmonic generation (HHG). (a) Schematic configuration for measurement of effect of photo-carrier doping on HHG. First, photo-carriers are generated in a solid by a near-infrared (NIR) pump pulse. After a long enough time passes for the photo-carriers to become incoherent, high harmonics are generated by a strong mid-infrared (MIR) driving field. (b) Schematic illustration of electron-hole dynamics in band structure. Intraband current may be enhanced due to acceleration of the doped photo-carriers in the MIR field (i). Interband polarization may be suppressed due to phase-space filling (ii) and excitation-induced dephasing (iii).

To implement such non-perturbative spectroscopies, a deeper understanding of the mechanism underlying HHG is crucial. A number of studies have treated light-driven coherent electrons and holes to describe the intraband and interband mechanisms [4–8,11,12,14,15]. The intense MIR pulses excite electron-hole pairs around the bandgap through a tunneling process. These electron-hole pairs are then driven to a higher energy region in k-space. Their nonlinear motion in real space, caused by anharmonicity in the band structure, induces intraband current and harmonic generation. The recombination of electron-hole pairs driven to a higher energy region also leads to harmonic generation through interband currents. In solids, many-body effects between electrons and holes, such as electron-electron scattering, may strongly affect the properties of HHG [5,7,15]. For example, the scattering of electron-hole pairs randomizes the phase of the interband polarization and the carrier distributions in k-space and creates incoherent carriers, as has been confirmed by the appearance of photoluminescence in HHG measurements [4,5]. However, achieving a deeper understanding of the HHG mechanisms, such as one disentangling the contributions of the intraband and interband processes, remains a challenge because of the inseparable carrier generation and acceleration processes due to the use of a single MIR pulse.

Here, a photo-carrier-doping experiment would be particularly useful for clarifying the HHG mechanisms [16]. Photo-carrier doping can be used to separately control the extent of carrier generation and thereby clarify the roles of the excited carriers. Figure 1(a) schematically shows a photo-carrier doping experiment. A near-infrared (NIR) pulse is applied to a sample to create electron-hole pairs. After a long enough time passes for the electron-hole pairs to become incoherent, a strong MIR pulse is applied to the sample to generate high harmonics. These photo-carriers add to the incoherent carriers excited by the MIR pulse, and they enhance or suppress the harmonic intensity (Fig. 1(b)). The photo-carrier doping may enhance the intraband current by increasing the total number of incoherent carriers driven by the strong field [17]. The magnitude of the intraband contribution in the HHG process can be estimated from the extent of harmonic intensity enhancement. In contrast, the photo-carriers may suppress the interband polarization since they disturb the carrier generation process in the MIR field (through the phase-space filling effect). This effect is deemed to be a major factor in the experimentally observed reduction in intensity of harmonics [16]. Photo-carrier doping may also promote excitation-induced dephasing (EID), which is a carrier-density-dependent electron-electron scattering effect, and also suppress the interband polarization [18]. This effect has been used to explain the broadening of the homogeneous linewidth of excitons in semiconductors [19–21]. We thus should be able to reveal the mechanisms underlying HHG by experimentally evaluating changes in harmonic intensity due to photo-carrier doping and systematically examining the above effects in a theoretical calculation. Such a systematic study has not been performed until now.

In this study, we conducted pump-probe experiments and numerical simulations to explore the effect of incoherent electron-hole pairs on HHG. The sample was a monolayer semiconducting transition metal dichalcogenide (TMD) [22]. This material is ideal for avoiding propagation effects, such as phase matching and reabsorption, which obscure the intrinsic harmonic spectrum and hinder the clarification of the microscopic HHG mechanisms [9,23–29]. The photo-carrier-doping experiment indicated an enhancement in harmonics at fifth order and reductions at seventh and higher order. The numerical calculations, incorporating the distribution of the photo-carriers, excitonic Coulomb interaction, and electron-electron scattering, systematically explored the effect of incoherent carrier doping. We found that the lower order harmonics below the bandgap are enhanced by the intraband acceleration of the incoherent photo-carriers, which is dampened by a momentum relaxation process. We also found that the origin of the suppression of the interband polarization above the bandgap is EID rather than the phase-space filling effect. Our numerical calculations suggest that incoherent carriers generated by a strong driving field should strongly reduce the efficiency of generating higher-order interband harmonics.

This paper is organized as follows. Section II describes our samples and the experimental setup. Section III presents the results of pump-probe spectroscopy experiments to estimate the photo-carrier density inside the sample against the pump fluence and pump-probe time delay. Section IV describes the experimental investigation of the photo-carrier doping effect on the harmonic generation. The method of the numerical simulation is presented in section V, and the results of the full calculation are shown in section VI. The effects of the initial carriers, momentum relaxation, EID and excitons on the photo-carrier doping are systematically discussed by using switch-off analyses in section VII. Section VIII discusses the contributions of incoherent carriers generated by the MIR driving field. Section IX discusses the screening effect of excitonic Coulomb interaction at high carrier densities.

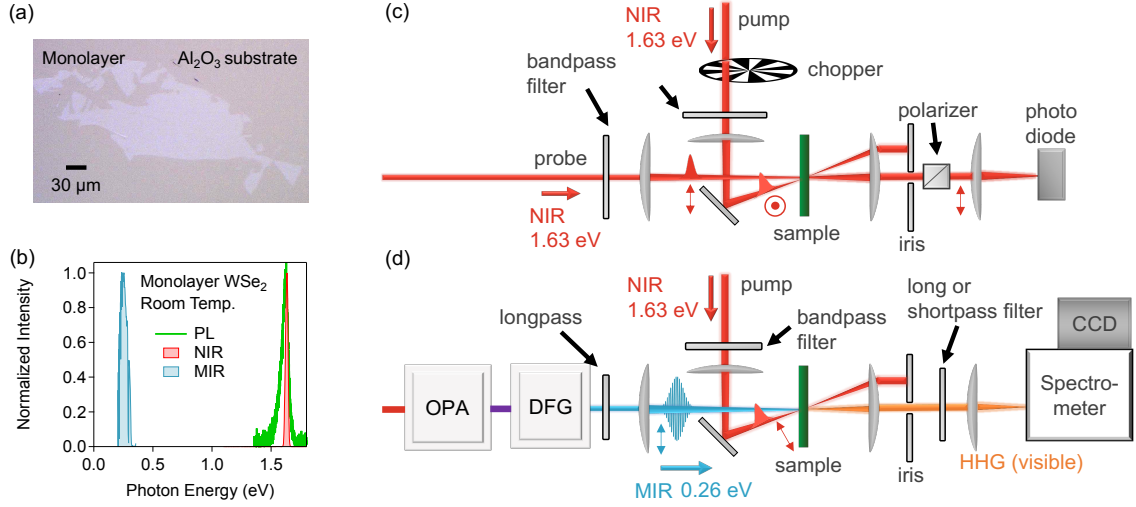

Fig. 2 Sample and experimental setup. (a) Photograph of WSe$_2$ monolayer sample on sapphire substrate. (b) Photoluminescence spectrum of monolayer WSe$_2$ and mid-infrared and near-infrared pulse spectra. (c) Experimental setup of degenerate near-infrared pump-probe measurement. The laser source is a Ti: sapphire regenerative amplifier (photon energy 1.55 eV, 35 fs pulse duration, 1 kHz repetition rate, 7 mJ pulse energy). The near-infrared (NIR) pump and probe beams pass through bandpass filters (1.63 eV) and are incident on the sample with polarizations perpendicular to each other. The transmittance of the probe beam is measured by a photo diode and read out by a lock-in amplifier. (d) Experimental setup for measuring the effect of photo-carrier doping on high harmonic generation (HHG). The NIR pump (1.63 eV) and strong mid-infrared (MIR) beams (0.26 eV) are incident on the sample with parallel polarizations. The harmonic spectra are resolved by a spectrometer and measured by a Si-CCD camera (OPA: optical parametric amplifier, DFG: difference frequency generation).

## II. Sample preparation and optical setup

We purchased a bulk WSe$_2$ crystal from 2D Semiconductors Inc. We fabricated an isolated monolayer WSe$_2$ sample on a sapphire substrate with a thickness of 0.43 mm by using the mechanical exfoliation and dry transfer method (Fig. 2(a)). The photoluminescence (PL) spectra were recorded with a commercial micro PL spectrometer (NanoFinder30, Tokyo Instruments Inc.) at room temperature. The PL spectra were used to determine the photon energy for the photo-carrier doping. A typical PL spectrum of the monolayer WSe$_2$ sample is shown in Fig. 2(b), which was obtained after the HHG measurement (PL spectra between before and after the HHG measurement are compared in Fig. S5). The PL peak is located at 1.63 eV, which corresponds to the A-exciton energy of the monolayer WSe$_2$ [22]. In the following, the A-exciton energy is defined as the absorption edge.

Figure 2(c) shows the optical setup for degenerate NIR pump-probe spectroscopy of the photo-carrier dynamics. Part of the output of a Ti: sapphire based regenerative amplifier (photon energy 1.55 eV, 35 fs pulse duration, 1 kHz repetition rate, 7 mJ pulse energy) was passed through a bandpass filter to choose the photon energy resonant with the A-exciton energy (1.63 eV, bandwidth: 10 nm). The spectrum of the NIR pulse is shown in Fig. 2(b). The pump and probe beams were set to have linear polarizations perpendicular to each other. The spot sizes of the pump and probe NIR pulses in the focal plane were respectively 60 μm and 30 μm, full width at half maxima (FWHM), assuming a Gaussian profile. The pump beam was dumped by an iris and a polarizer after it went through the sample. The transmitted probe beam was detected by a photo-detector and read out by a lock-in amplifier tuned to a chopper frequency of 500 Hz.

Figure 2(d) shows the setup for the optical pump-HHG probe measurement to examine the effect of photo-carrier doping on the HHG. The setup is basically the same as the one in Ref. [29]. The same regenerative amplifier was used to pump the optical parametric amplifier (TOPAS-C, Light Conversion). Strong MIR pulses (photon energy 0.26 eV) were obtained by difference frequency generation (DFG) of the signal and idler beams in a $AgGaS_2$ crystal. The signal and idler beams were blocked by a long pass filter (LPF) with a cutoff wavelength of 4 μm. The spectrum of the MIR pulse is shown in Fig. 2(b). The photon energy was chosen far below the resonance with the absorption edge of the monolayer $WSe_2$ to avoid damaging the sample in the strong field. The pulse duration was estimated to be 60 fs in a cross-correlation measurement [29]. The same NIR pump pulses as in Fig. 2(c) were used for the photo-carrier doping. The polarizations of NIR and MIR pulses were parallel to each other. The NIR light was spatially separated and blocked by an iris placed behind the sample. The NIR and MIR pulse spot sizes were 60 μm and 30 μm (FWHM), respectively. High harmonics were spectrally resolved with a grating spectrometer (Kymera 193i, Andor) and measured with a Si charge-coupled device camera (DU920P-OE Newton). The scattered component of the NIR light was blocked by a 750-nm short pass filter (SPF) when measuring the seventh and higher order harmonics or by a 850-nm LPF when measuring the fifth-order harmonics. All the degenerate pump-probe measurements and HHG measurements were performed in air at room temperature.

**III. Degenerate NIR pump-probe measurement**

We performed a degenerate NIR pump-probe experiment to estimate the photo-carrier density for the excitation photon energy resonant with the A-exciton energy. Figure 3 shows the dependence of the differential transmission signal on the time delay. The differential transmission ($\Delta T$) is normalized by the transmission of the probe pulse without the pump pulse ($T$). The inset in Fig. 3 shows the overall measured data at four excitation fluences of 40, 80, 160, 320 μJ cm$^{-2}$. Increases in transmission ($\Delta T/T > 0$) were obtained in all experiments. This indicates that the absorption saturation of the probe pulse was due to the phase-space filling effect. The spike-like signal observed near the time origin may reflect a coherent nonlinear process caused by the pump and probe pulses. Thus, we evaluated only the decay curves after the spikes shown in the main panel of Fig. 3. The decay rate of $\Delta T/T$ grows with increasing pump fluence.

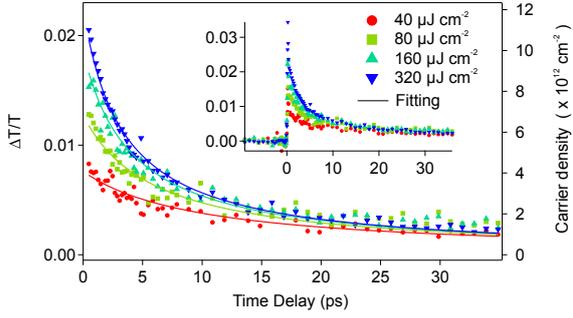

Fig. 3 Photo-carrier relaxation dynamics at room temperature. Dependence of normalized differential transmission on time delay for four excitation fluences (40, 80, 160, 320 μJ cm$^{-2}$). Solid lines are fitting curves assuming the exciton-exciton annihilation (EEA) process and absorption saturation of the pump pulse. Inset: full-range data. The fitting result of the EEA rate is $k_A = (2.46 \pm 0.10) \times 10^{-2}$ cm$^2$ s$^{-1}$ and the saturation fluence is $F_S = (1.014 \pm 0.039) \times 10^2$ μJ cm$^{-2}$. The right axis represents the corresponding carrier density. The carrier density of $1.1 \times 10^{13}$ cm$^{-2}$ is close to the carrier number of 0.01 per unit cell.

This indicates that the relaxation process becomes faster nonlinearly with respect to the carrier density. The exciton-exciton annihilation (EEA) process occurs in TMDs at around an excitation fluence 40 μJ cm$^{-2}$, as reported in Refs. [30,31]. In addition, $\Delta T/T$ do not increase proportionally to the pump fluence, which indicates saturation of the carrier generation by the pump pulse. By considering these two effects, we fitted the data with the following function:

$$\frac{\Delta T}{T} = \frac{\hbar \omega_N}{F_S} \frac{2}{n_S + 1} N(t), \tag{1}$$

where

$$N(t) = \frac{N_0}{1 + k_A N_0 t} \tag{2}$$

is the carrier (exciton) density in the 2D sample plane at the time delay $t$ assuming the EEA process and

$$N_0 = \frac{A F_S}{\hbar \omega_N}\left(1 - \exp\left(-\frac{F_N}{F_S}\right)\right) \tag{3}$$

is the carrier density at the time origin considering the saturation of the carrier generation by the pump pulse (For a detailed derivation, see Supplementary Material). Here, $\hbar \omega_N = 1.63$ eV is the photon energy of the NIR pulse, $F_S$ is the saturation fluence, $F_N$ is the pump fluence, $k_A$ is the EEA rate, $A$ is the absorbance of the monolayer WSe$_2$, and $n_S = 1.76$ is the refractive index of the sapphire substrate [32]. Multiple reflections in our optical system are neglected in these equations. A global fitting using Eqs. (1)-(3) was performed on the four measured data with $F_S, k_A, A$ as common free parameters. The fitting results are displayed in Fig. 3 as solid curves. The fitting curves successfully explain all the decay curves. The fitting results are summarized in Table S1. The rate $k_A = (2.46 \pm 0.10) \times 10^{-2}$ cm$^2$ s$^{-1}$ is close to the values reported in Refs. [30,31]. Furthermore, this result validates the estimation of the photo-carrier density. The right axis in Fig. 3 shows the carrier density estimated using Eq. (1) and the fitting result of $F_S = (1.014 \pm 0.039) \times 10^2$ μJ cm$^{-2}$. The estimated carrier density in Fig. 3 is close to, but smaller than, the value at which the bandgap renormalization becomes remarkable ($N \sim 10^{13} - 10^{14}$ cm$^{-2}$) [33]. In the next section, we will mainly examine the results on photo-carrier doping at a pump fluence of 40 μJ cm$^{-2}$ (corresponding to the red circle in Fig. 3) to minimize the impact of the bandgap renormalization.

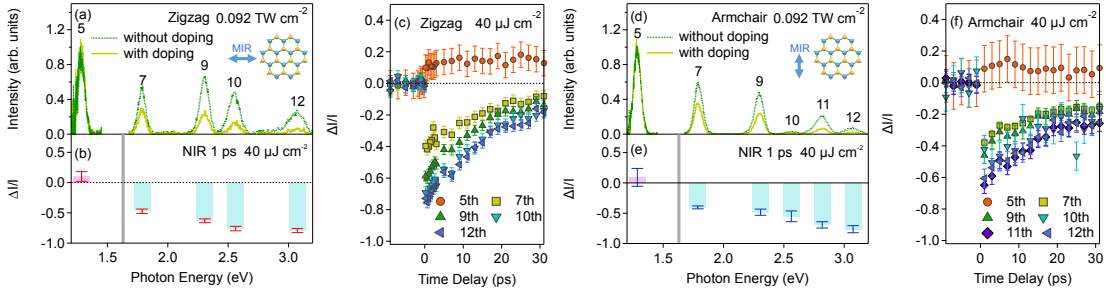

Fig. 4 Effect of photo-carrier doping on high harmonic generation in monolayer WSe$_2$ at room temperature. (a,d) Harmonic spectra induced by intense mid-infrared (MIR) pulses polarized along (a) zigzag and (d) armchair directions. Green dotted and yellow solid spectra were respectively obtained without and with near-infrared (NIR) pump pulses. The inset is a top view of the crystal structure of monolayer WSe$_2$ and the light blue arrow represents the polarization of the MIR pulses. (b,e) NIR-pump-induced differential harmonic intensity for the pulses polarized along (b) zigzag and (e) armchair directions. The time delay and the NIR pump fluence were set to 1 ps and 40 μJ cm$^{-2}$, respectively. (c,f) Dependence of the differential harmonic intensity on time delay for the pulses polarized along (c) zigzag and (f) armchair directions with pump fluence of 40 μJ cm$^{-2}$. The error bars represent standard deviations calculated by assuming that the deviations are independent of the time delay. The gray lines in (b) and (e) show the A-exciton energy where the photo-carriers are excited by the NIR pulses.

**IV. Effect of photo-carrier doping on HHG**

Figure 4(a) shows the fifth to twelfth HHG spectra with and without photo-carrier doping. The polarization of the driving MIR pulses was along the zigzag direction (inset of Fig. 4(a)). The crystal axis was determined according to the polarization selection rules of HHG [25,26]. The peak intensity of the MIR pulse was 0.092 TW cm$^{-2}$ inside the sample, which was calculated by considering the incident pulse intensity, spot sizes, pulse duration and refractive index of the sapphire substrate [32]. Even-order harmonics appeared because of the broken inversion symmetry of the monolayer [25,26]. The time delay was set to 1 ps to prevent consecutive pulses from overlapping, where the positive time delay means that the NIR pump pulses arrive before the MIR pulses. The pump fluence was set to 40 μJ cm$^{-2}$, which corresponds to a photo-carrier density of $3.7 \times 10^{12}$ cm$^{-2}$. We observed a clear difference in harmonic intensity induced by the pump pulse. Figure 4(b) shows the differential harmonic intensity normalized by the harmonic intensity for each order ($\Delta I/I$), where $I$ is the intensity without photo-carrier doping. The intensity $I$ is calculated by averaging the harmonic spectra over their spectral widths (FWHM). In the energy region below the absorption edge (1.63 eV), we observed a small increase in the intensity of the fifth-order harmonics. In contrast, the higher order harmonics are clearly suppressed, and the degree of suppression becomes larger as the order increases. We found no clear difference between odd and even harmonics. Figures 4(d) and 4(e) show the results for the measurement with the pulses polarized along the armchair direction (see Fig. 4(d) inset). The differential harmonic intensity in Fig. 4(d) shows a trend similar to the one for the zigzag direction. These results suggest that the differential harmonic intensity is determined solely by the energy of the harmonics relative to that of the absorption edge.

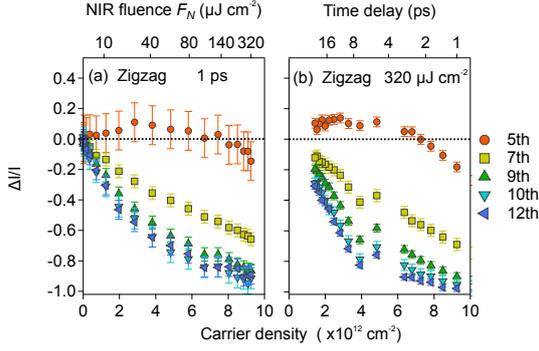

Fig. 5 Photo-carrier density dependence of differential harmonic intensity. (a) The dependence of the near-infrared (NIR) pump fluence at a time delay of 1 ps and (b) the dependence on the time delay at a pump fluence of 320 μJ cm$^{-2}$ are converted into carrier density dependences. The carrier density was estimated from the fitting results in Fig. 3. The mid-infrared polarization is along the zigzag direction. The error bars represent standard deviations calculated by assuming that the deviations are independent of the time delay and NIR pump fluence. The small dip near 7 ps in (b) may have been caused by multiple reflections from our optical system.

To confirm the relation between the HHG and photo-carrier dynamics, we measured the differential harmonic intensity as a function of the time delay using the pulses polarized along the zigzag and armchair directions at 40 μJ cm$^{-2}$ (Figs. 4(c) and 4(f)). In both experiments, the signal appeared as a staircase around the time origin, and it decreased on a time scale of about 10 ps. This timescale matches the one of the photo-carrier dynamics in Fig. 3.

To evaluate the dependence of the differential harmonic intensity on the photo-carrier density, we converted the NIR fluence dependence and the time delay dependence by using the fitting results of Eqs. (1-3). The dependences in Figs. 5(a) and 5(b) were obtained at a delay of 1 ps and at an NIR pump fluence of 320 μJ cm$^{-2}$, respectively. The figures indicate that the observed differential HHG signal is determined only by the carrier density at the time of arrival of the MIR pulses. This suggests that other photo-excitation effects, such as temperature changes in the electron and phonon system, are negligible for HHG. We observed a monotonic decrease for the higher order harmonics but a non-monotonic trend for the fifth-order harmonics at higher pump fluences.

Since the MIR field itself creates electron-hole pairs, the effects of the photo-carriers generally depend on the MIR intensity and underlying HHG mechanism. We measured the MIR intensity dependence of the differential harmonic intensity by using pulses polarized along the zigzag direction, as shown in Fig. 6. The photo-carrier density was estimated to be $3.7 \times 10^{12}$ cm$^{-2}$ under a pump fluence of 40 µJ cm$^{-2}$ and a delay of 1 ps. There was no significant change in differential harmonic intensity with respect to the MIR intensity, especially for the seventh and higher harmonics. The increase in the fifth-order harmonics was relatively large at lower MIR intensities. The signal-to-noise ratio became worse at intensities below 0.04 TW cm$^{-2}$. These experimental data will be compared with the numerical simulations in sections VII and VIII.

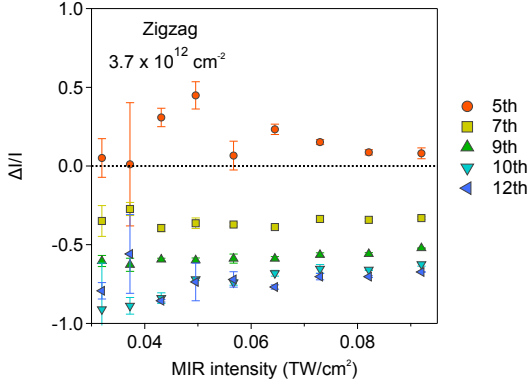

Fig. 6 Mid-infrared (MIR) intensity dependence of differential harmonic intensity under photo-carrier doping. The polarization of the MIR pulse is along the zigzag direction. The error bars represent standard deviations from the mean of five measurements under identical conditions.

We concluded that the observed decreases in harmonic intensity were not due to screening of the MIR field by the photo-carriers. We measured the transmittance of the MIR pulses under the same conditions as those of the experiment in Fig. 4(b). The transmission loss due to the photo-carrier doping was less than 1 %. This effect is not large enough to reduce the intensity of harmonics, given that the harmonic intensity is proportional to the cube of the MIR intensity [25,26].

## V. Method of numerical calculation

To better understand the effect of photo-carrier doping, we theoretically calculated harmonic spectra by using the semiconductor Bloch equations (SBEs) [34]. These equations are based on the single-particle picture within the Hartree-Fock approximation, and the Coulomb interaction is introduced between the individual electrons and holes. To solve the SBEs, we used the band model of gapped graphene [35]. This model describes the band structure near the band edge of monolayer TMDs at the K and K' points [36]. The energy difference between the conduction and valence bands are assumed to be

$$\varepsilon_g(\mathbf{k}) = \sqrt{\Delta^2 + 4t_r^2|f(\mathbf{k})|^2}. \qquad (4)$$

Here, $\Delta = 1.89$ eV is the bandgap at the K point and

$$f(\mathbf{k}) = \sqrt{1 + 4\cos\frac{\sqrt{3}k_x a}{2}\cos\frac{k_x a}{2} + 4\cos^2\frac{k_x a}{2}}, \qquad (5)$$

where $a = 3.28$Å is the lattice constant of monolayer WSe$_2$ taken from experimental results [37,38]. The hopping integral $t_r = 1.47$ eV is determined so that the effective mass of the conduction band electron at the K point matches the value estimated from a first principles calculation [39]. The band structure is shown in Fig. 7(a). In our calculation, the dipole moment between the conduction and valence bands $\mathbf{d}_{cv}(\mathbf{k})$ is assumed to be constant in k-space for simplicity. The Berry connection is also neglected for simplicity. The Coulomb interaction matrix $V_\mathbf{K}$, which is in general dependent on the wavenumbers of individual electrons [40], is assumed to be only dependent on the relative wavevector of the electrons. Under these conditions, the system recovers inversion symmetry, which results in the disappearance of the even-order harmonics in the simulation. This is not an issue in our simulation as there was little difference between the behaviors of the odd- and even-order harmonics in our experiment. The SBEs are written as

$$\frac{\partial}{\partial t}\tilde{P}_\mathbf{K} = -\frac{i}{\hbar}\left[e_g\left(\mathbf{K}+\frac{e}{\hbar}\mathbf{A}(t)\right) - i\gamma - i\gamma_e N\right]\tilde{P}_\mathbf{K} - i\omega_{R,\mathbf{K}+\frac{e}{\hbar}\mathbf{A}(t)}(2\tilde{n}_\mathbf{K}-1), \quad (6)$$

$$\frac{\partial}{\partial t}\tilde{n}_\mathbf{K} = -2\mathrm{Im}\left[\omega_{R,\mathbf{K}+\frac{e}{\hbar}\mathbf{A}(t)}\tilde{P}_\mathbf{K}^*\right] - \frac{1}{\hbar}\gamma_m\left[\tilde{n}_\mathbf{K} - n_{\mathbf{K}+\frac{e}{\hbar}\mathbf{A}(t),eq}\right], \quad (7)$$

where $\tilde{P}_\mathbf{K}$ and $\tilde{n}_\mathbf{K}$ are the interband coherence and population, respectively (see the Supplementary Material for the detailed derivation). The renormalized single-particle bandgap energy $e_g(\mathbf{K})$ and the generalized Rabi frequency $\omega_{R,\mathbf{K}}$ are written as

$$e_g\left(\mathbf{K}+\frac{e}{\hbar}\mathbf{A}(t)\right) = \varepsilon_g\left(\mathbf{K}+\frac{e}{\hbar}\mathbf{A}(t)\right) - 2\sum_{\mathbf{Q}\neq\mathbf{K}} V_{\mathbf{K}-\mathbf{Q}}\,\tilde{n}_\mathbf{Q}, \quad (8)$$

$$\omega_{R,\mathbf{K}+\frac{e}{\hbar}\mathbf{A}(t)} = \frac{1}{\hbar}\left[\mathbf{d}_{cv}\cdot\mathbf{E}(t) + \sum_{\mathbf{Q}\neq\mathbf{K}} V_{\mathbf{K}-\mathbf{Q}}\,\tilde{P}_\mathbf{Q}\right], \quad (9)$$

respectively [34]. The momentum $\hbar\mathbf{k} = \hbar\mathbf{K} + e\mathbf{A}(t)$ describes the center of momentum of the electron wave packet following the acceleration theorem, where $\hbar\mathbf{k}$ is the crystal momentum and $e > 0$ is the elementary charge. Here, a symmetric population in the conduction and valence bands $\tilde{n}_\mathbf{K} = \tilde{n}_{c,\mathbf{K}} = 1 - \tilde{n}_{v,\mathbf{K}}$ is assumed. We computed the electric field of the MIR driving field $\mathbf{E}(t) = -\partial\mathbf{A}(t)/\partial t$, which had a pulse width of 60 fs and a peak field strength of 8.3 MV cm$^{-1}$ corresponding to the intensity in our experiment, 0.092 TW cm$^{-2}$. The excursion scale of the electron-hole pairs in k-space is about a quarter of the distance between the K and K' points at this electric field. The MIR polarization is fixed to be only along the zigzag direction in Fig. 7(c), because the experiment showed no clear difference between the zigzag and armchair polarizations. The phenomenological dephasing rate $\gamma$ of the interband polarization takes into account the electron-phonon scattering. The spontaneous recombination and Auger recombination process of the electron-hole pairs are assumed to be slow enough to be neglected in the HHG process. The EID coefficient $\gamma_e$ due to the electron-electron scattering is assumed to increase linearly with the total carrier density $N$ [18–21]. The total dephasing rate is defined as $\Gamma = \gamma + \gamma_e N$. The carrier density $N$ was estimated by dividing the carrier number $\sum_\mathbf{Q} \tilde{n}_\mathbf{Q}$ by the system size, which is the area of the unit cell of monolayer WSe2 multiplied by the number of k-mesh points.

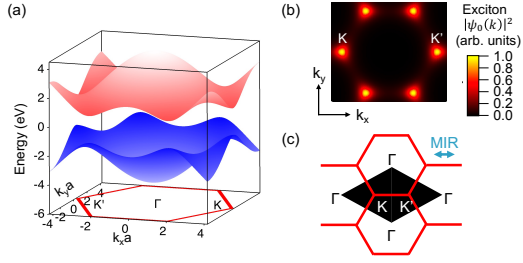

Fig. 7 Method of numerical calculation (a) Band structure of gapped graphene model (red: conduction band, blue: valence band). (b) Lowest energy exciton wavefunction squared $|\psi_0(k)|^2$ distributed in k-space in the single electron-hole picture. The exciton binding energy was set to 0.26 eV. (c) Calculated area (black) in k-space that corresponds to the first Brillouin zone. The light blue arrow represents the mid-infrared polarization (zigzag) used in the calculation.

The photo-carriers are included in our simulation as initial carriers before the MIR irradiation. In our experiment, the NIR pulses were resonant with the A-exciton peak. Thus, the photo-carrier distribution is broadened in k-space according to the exciton wave function [22]. The exciton wave function is calculated by diagonalizing the following eigenvalue problem:

$$\left(\varepsilon_g(\mathbf{k}) - \varepsilon_\nu\right)\psi_\nu(\mathbf{k}) = \sum_{\mathbf{q} \neq \mathbf{k}} V_{\mathbf{k}-\mathbf{q}} \psi_\nu(\mathbf{q}), \tag{10}$$

where $\psi_\nu(\mathbf{k})$ is the exciton wave function in k-space in the single-particle picture and $\varepsilon_\nu$ is the corresponding exciton energy [34]. The Coulomb interaction is modeled with the Rytova-Keldysh potential, $V_\mathbf{q} = \sum_G V_{RK,\mathbf{q}+\mathbf{G}}$, where

$$V_{RK,\mathbf{q}} = \frac{e^2}{2\epsilon|\mathbf{q}|(1+\rho_0|\mathbf{q}|)}, \tag{11}$$

where the screening length $\rho_0 = 0.95$ nm [41,42]. The sum over the reciprocal vectors $\mathbf{G}$ is calculated up to the third-nearest Brillouin zone to ensure translational symmetry of $V_\mathbf{q}$. The average dielectric constant of the surrounding media was chosen to be $\epsilon = 3.98\epsilon_0$ so that the lowest energy of the exciton corresponds to the A-exciton energy of the monolayer WSe2 in our experiment (1.63 eV), where $\epsilon_0$ is the vacuum permittivity. The photo-carrier distribution in k-space is assumed to be proportional to the square of the lowest energy exciton wave function $|\psi_0(\mathbf{k})|^2$ (Fig. 7(b)). We prepared $|\psi_0(\mathbf{k})|^2$ by using a fitting function of the numerically calculated value with the following trial function:

$$|\psi_0(\mathbf{k})|^2 = \left(\frac{c_0}{(c_1 + |f(\mathbf{k})|^2)^{c_2}} + c_3\right)\exp\left(-\frac{|f(\mathbf{k})|^2}{\sigma^2}\right). \tag{12}$$

The fitting parameters are shown in Table S2. The maximum residual of the fitting was 0.9% of the maximum value, indicating that the trial function reproduced the numerically obtained wave function.

We phenomenologically treat the momentum relaxation as a relaxation into the equilibrium distribution [43]. We use $\gamma_m$ in Eq. (7) to describe the momentum relaxation rate, taking into account the electron-electron and electron-phonon scattering. We use the equilibrium carrier distribution $n_{\mathbf{k},eq}$ defined by the exciton wave function and total carrier density:

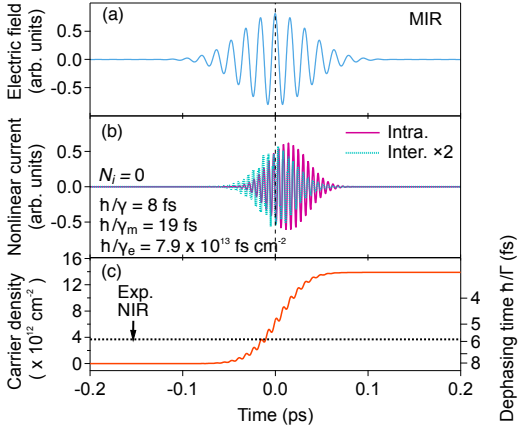

Fig. 8 Calculated intraband and interband current and evolution of carrier density in the time domain. (a) Time profile of the mid-infrared (MIR) field used in the calculation. (b) Nonlinear intraband and interband current calculated without initial carriers ($N_i = 0$). The interband current is multiplied by 2 for visibility. (c) Time evolution of carrier density and corresponding dephasing time $\hbar/\Gamma$. The dotted black line represents the photo-carrier density of $3.7 \times 10^{12}$ cm$^{-2}$ excited by the near-infrared (NIR) pump pulse, corresponding to the experimental condition in Fig. 4. The calculation parameters are $\hbar/\gamma = 8$ fs, $\hbar/\gamma_m = 19$ fs, and $\hbar/\gamma_e = 7.9 \times 10^{13}$ fs cm$^{-2}$. The dephasing time is $\hbar/\Gamma = 3.3$ fs after the MIR pulse passes.

$$n_{\mathbf{k},eq} = \frac{|\psi_0(\mathbf{k})|^2}{\sum_q |\psi_0(\mathbf{q})|^2} N \qquad (13)$$

so that the total carrier density is conserved when the interband transition is zero. Equations (6,7) are calculated using a 75×75 k-mesh in the region shown in Fig. 7(c), which is equivalent to the first Brillouin zone. We estimated the magnitude of the dipole moment of the band-to-band transition $d_{cv}$ from the magnitude of the excitonic optical Stark shift [44]. The magnitude of the excitonic optical Stark shift is enhanced relative to the two-level optical Stark shift by a factor $\rho_{1S} = \psi_{1S}(\mathbf{r} = 0)\sum_\mathbf{k}\psi_{1S}(\mathbf{k})|\psi_{1S}(\mathbf{k})|^2$ [34]. Thus, we estimated $d_{cv} = 1.5 \times 10^{-29}$ [C·m] by dividing the dipole moment of the exciton by the square root of the enhancement factor.

To calculate the harmonic spectra, we evaluated the harmonic intensity from the intraband and interband currents defined as [14,45]

$$\mathbf{J}(t) = \mathbf{J}_{ra}(t) + \mathbf{J}_{er}(t) \qquad (14)$$

$$\mathbf{J}_{ra}(t) = \frac{2e}{\hbar}\sum_\mathbf{K} \frac{\partial}{\partial \mathbf{K}}\varepsilon_g\left(\mathbf{K} + \frac{e}{\hbar}\mathbf{A}(t)\right)\tilde{n}_\mathbf{K} \qquad (15)$$

$$\mathbf{J}_{er}(t) = \frac{2}{\hbar}\sum_\mathbf{K} \varepsilon_g\left(\mathbf{K} + \frac{e}{\hbar}\mathbf{A}(t)\right)\text{Im}[\mathbf{d}_{cv}^*\tilde{P}_\mathbf{K}] \qquad (16)$$

We assumed that the harmonic spectra are proportional to $|\mathbf{J}(\omega)|^2$. Below, the harmonic intensity at each order is calculated by summing the spectra over the full harmonic spectral width.

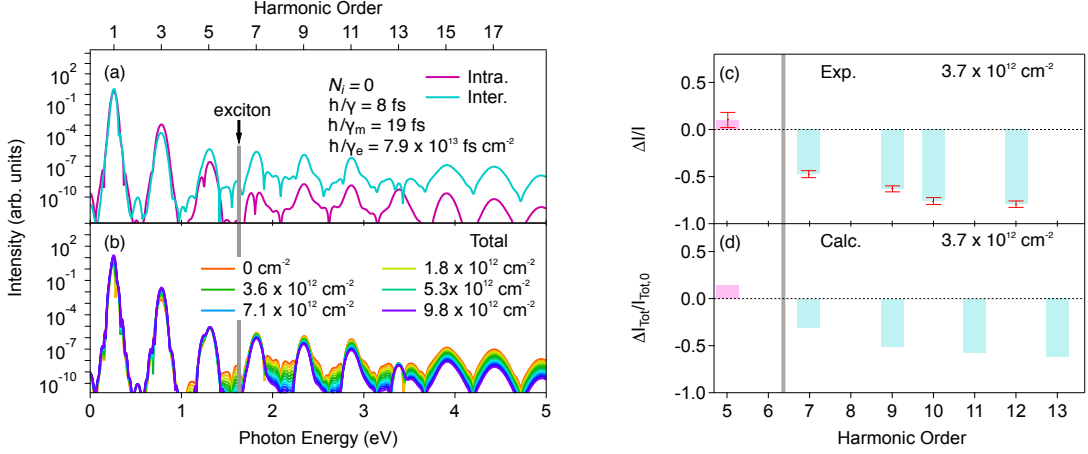

Fig. 9 Numerical calculation of high harmonic generation spectra with photo-carrier doping effects. The calculation parameters are $\hbar/\gamma = 8$ fs, $\hbar/\gamma_m = 19$ fs, and $\hbar/\gamma_e = 7.9 \times 10^{13}$ fs cm$^{-2}$. (a) Intraband and interband harmonic spectrum calculated without initial carriers ($N_i = 0$). (b) Total harmonic spectra for various photo-carrier densities. (c,d) Comparison of differential harmonic intensities between (c) experiment and (d) calculation at the initial carrier density of $3.7 \times 10^{12}$ cm$^{-2}$. The calculated result in (d) is the differential total harmonic intensity normalized by the total harmonics without initial carriers. The gray lines represent the lowest exciton energy.

## VI. Results of the numerical calculation

We solved the SBEs in Eqs. (6) - (9) and calculated the intraband and interband currents. We set the dephasing time $\hbar/\gamma$ to 8 fs (a half cycle of the MIR field), which is similar to the value in Ref. [46]. The momentum relaxation time $\hbar/\gamma_m$ was set to 19 fs, which is longer than the dephasing time $\hbar/\gamma$. The EID coefficient $\hbar/\gamma_e$ was set to $7.9 \times 10^{13}$ fs cm$^{-2}$ to reproduce the experimental results. The time evolution of the mid-infrared field and the intraband and interband currents without initial carriers are shown in Figs. 8(a) and 8(b). The time evolution of the carrier density is shown in Fig. 8(c). The carriers are generated by the tunneling process most efficiently when the electric field is at its peak. This process is repeated for half the period of the MIR electric field. Without dephasing, no real carrier excitation occurs under the non-resonant driving field. However, excited carriers accumulate when fast dephasing processes are present during the period of the MIR field [47]. Such incoherent carriers contribute to the intraband current by being accelerated within the anharmonic band structure; thus, the peak amplitude of the intraband current appears later in the pulse duration, as shown in Fig. 8(b) [15]. The calculated intraband and interband harmonic spectra without initial carriers are shown in Fig. 9(a). Here, only odd-order harmonics appear because the two-band system in the calculation has inversion symmetry. In this calculation, the interband harmonics are dominant in the energy region above the absorption edge (1.63 eV), whereas the intraband harmonics are relevant in the lower energy region, as several studies have indicated [15,17]. The other intraband contribution originating from the temporally oscillating population is a minor source of above-bandgap harmonics (see Fig. S1).

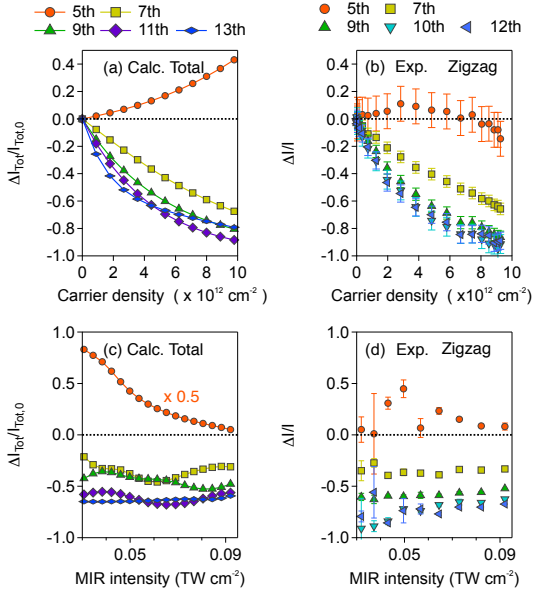

Fig. 10 Comparison of photo-carrier density dependence and mid-infrared (MIR) intensity dependence of differential harmonic intensity between calculation and experiment. The calculation parameters are $\hbar/\gamma = 8$ fs, $\hbar/\gamma_m = 19$ fs, and $\hbar/\gamma_e = 7.9 \times 10^{13}$ fs cm$^{-2}$. (a,b) Comparison of photo-carrier density dependence of differential harmonic intensity between (a) calculation and (b) experiment in Fig. 5 (a). (c,d) Comparison of MIR intensity dependence of differential harmonic intensity between (c) calculation at an initial carrier density of $3.7 \times 10^{12}$ cm$^{-2}$ and (d) experiment in Fig. 6. The fifth-order harmonic intensity in (c) is multiplied by 0.5 for visibility.

Next, we calculated the harmonics generated with a finite number of initial carriers. Figure 9(b) shows the total harmonic spectra for various photo-carrier densities. Figures 9(c) and 9(d) compare the experimental results with the numerical calculation at an initial carrier density of $3.7 \times 10^{12}$ cm$^{-2}$. The numerical calculation reproduced the behavior of the positive and negative changes below and above the absorption edge. There was also good agreement on the larger reduction of the higher-order harmonics.

The initial-carrier-density dependence of the differential total harmonic intensity was calculated as shown in Fig. 10(a). The corresponding experimental result is in Fig. 10(b). The photo-carrier density dependences are in good agreement especially for the higher-order harmonics, as is the magnitude of the increase at fifth order at lower initial carrier density. At higher carrier density, the calculated dependence for the fifth-order harmonics deviate from the dependence obtained from the experiment. In addition, we compared the calculated and experimental MIR intensity dependences at an initial carrier density of $3.7 \times 10^{12}$ cm$^{-2}$. The calculated differential harmonic intensity at seventh or a higher order in Fig. 10(c) does not significantly depend on the MIR intensity, which is consistent with the experimental results in Fig. 10(d). The trend, in which the differential harmonic intensity increases as the MIR intensity decreases, is the same as in the experiment. These results are discussed in sections VII and VIII.

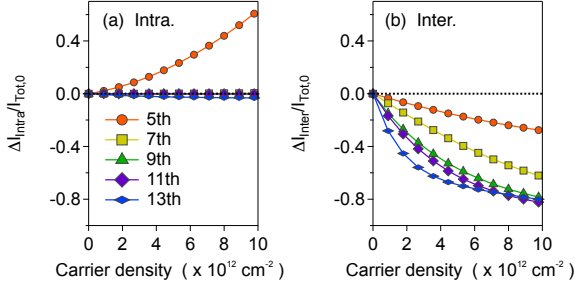

Fig. 11 Numerical calculation of photo-carrier density dependence of (a) intraband and (b) interband contributions to differential harmonic intensity. The calculation parameters are $\hbar/\gamma = 8$ fs, $\hbar/\gamma_m = 19$ fs, and $\hbar/\gamma_e = 7.9 \times 10^{13}$ fs cm$^{-2}$. Differential intraband and interband harmonic intensities are normalized by the total harmonic intensity calculated without initial carriers.

To understand the mechanism underlying the observed changes in harmonic intensity due to photo-carrier doping, we calculated the initial-carrier-density dependence of the intraband and interband currents. In Fig. 11, the differential harmonic intensity is normalized by the total harmonic intensity. These results indicate that the increase in the fifth-order harmonics is caused by an increase in the intraband contribution (Fig. 11(a)). On the other hand, the reductions in the harmonics at energies higher than the absorption edge can be attributed to the reduction in the interband contribution (Fig. 11(b)).

The increase due to the intraband contribution indicates the importance of considering photo-carriers in numerical simulations. Doping of photo-carriers can enhance the intraband current through the acceleration of the photo-carriers and also suppress the interband polarization through the phase-space filling effect. On the other hand, Heide *et al.* [18] did not find an increase in harmonics in their calculation. They did not include the photo-carrier distribution and only included the photo-carrier density dependence in the dephasing constant. To illustrate the importance of the photo-carrier distribution, we also performed a calculation without the initial carrier distribution (Fig. 12), similar to Ref. [18]. In this calculation, the EID coefficient $\gamma_e$ was set to 0 and the dephasing constant was effectively varied by amounts corresponding to the EID due to the pump pulse. Here, we could not reproduce the increase in the fifth-order harmonics. These results indicate that the intraband acceleration of the photo-carriers is needed to explain the increase in intensity of the fifth-order harmonics, which is consistent with the results in Fig. 11(a). On the other hand, we found little difference in the above-bandgap harmonics. This result indicates that intraband acceleration of the incoherent carriers does not affect the harmonics above the absorption edge, where the interband mechanism is dominant.

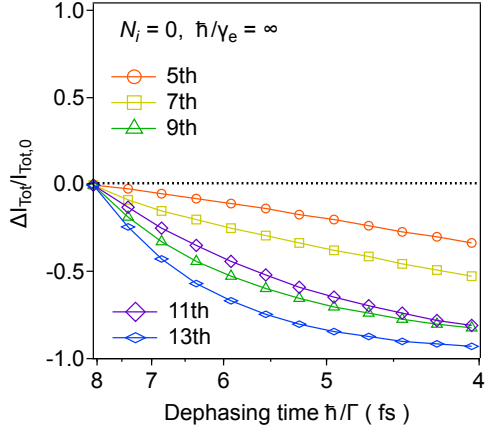

Fig. 12. Numerical calculation of differential harmonic intensity due to the change in dephasing time induced by the pump pulse. The calculations were performed without initial carriers ($N_i = 0$) by setting the dephasing times $\hbar/\Gamma$ constant in the time evolution ($\hbar/\gamma_e = \infty$). The calculated differential harmonic intensity is normalized by the total harmonic intensity at the dephasing time $\hbar/\Gamma = 8$ fs.

**VII. Switch-off analysis of the photo-carrier doping effects**

To clarify the many-body effects included in our calculation, we performed switch-off analyses based on the full numerical simulations described in the previous section. Figure 13(a) is a schematic diagram of the three effects included in our calculation: (ε1) momentum relaxation, (ε2) excitation-induced dephasing (EID), and (ε3) excitonic effect. Here, we discuss the initial-carrier-density dependence of the differential harmonic intensity on the three effects by removing their contributions one at a time.

Figure 13(b) shows the results for momentum relaxation. Without momentum relaxation ($\hbar/\gamma_m = \infty$), the fifth-order harmonics increase by one order of magnitude and clearly deviate from the experimental observations, where the enhancement was only about 10%. This is because the intraband current is suppressed by the momentum relaxation, as shown in Fig. S2. This indicates the significance of the momentum relaxation in the HHG process.

Figure 13(c) shows the results for the EID. Without EID ($\hbar/\gamma_e = \infty$), we could not reproduce the large suppression in higher order harmonics above the absorption edge, where the interband mechanism is dominant. Given what is shown in Fig. 12, this result indicates that the EID should be the main cause of the reduction in HHG, not the phase-space filling effect.

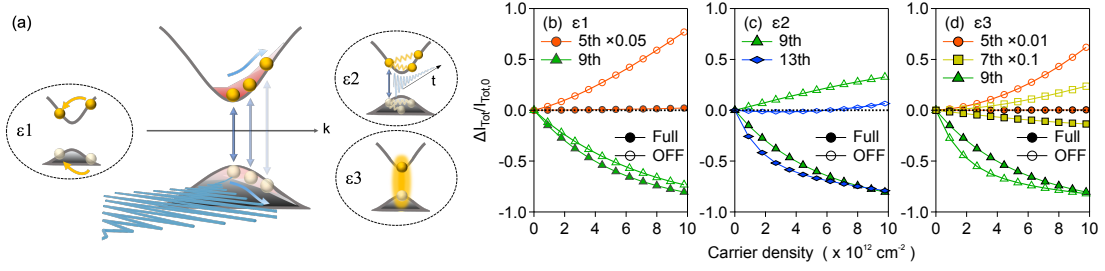

Fig. 13. Switch-off analysis of many-body effects: (ε1) momentum relaxation (ε2) excitation-induced dephasing (EID), and (ε3) excitonic effect. (a) Schematic illustration of the three effects. (b-d) Calculated initial-carrier-density dependence of differential total harmonic intensity normalized by the total harmonic intensity without initial carriers. Filled and unfilled symbols represent the results of full and switched-off calculations, respectively. The switched-off results in (b) were calculated by respectively setting $\hbar/\gamma_m = \infty$, (c) $\hbar/\gamma_e = \infty$, and (d) $V_q = 0$. The fifth-order harmonic intensity in (b) is multiplied by 0.05, and the fifth-order and seventh-order harmonic intensities in (d) are multiplied by 0.01 and 0.1, respectively, for visibility.

Figure 13(d) shows results for the excitonic effect derived from the strong excitonic Coulomb interaction in monolayer WSe$_2$. Without the excitonic effect ($V_q = 0$), there are large increases in the fifth- and seventh-order harmonics. If excitonic effects were absent, the bandgap would be between the energies of the seventh and ninth harmonics (1.89 eV). Thus, the intraband contribution would be relatively large so that the seventh-order harmonic intensity increases with the initial carrier density (yellow unfilled squares in Fig. 13(d)). This indicates that the excitonic absorption edge and not the single-electron bandgap energy determines the HHG mechanisms near the edge.

The above discussion makes it clear that we need to include all three contributions, i.e., the momentum relaxation, excitation-induced dephasing (EID), and excitonic effect, to reproduce the experimental effects of photo-carrier doping on HHG. Note that the EID coefficient $\hbar/\gamma_e = 7.9 \times 10^{13}$ fs cm$^{-2}$, i.e., $\gamma_e = 8.4 \times 10^{-12}$ meV cm$^2$, with which our calculation reproduces the experimental results, is of the same order of magnitude as the value, $\gamma_e = 2.7 \times 10^{-12}$ meV cm$^2$, estimated from the homogeneous linewidth of the exciton in monolayer WSe$_2$ at low temperature [21]. This result points to the validity of including EID in our calculation.

**VIII. Contributions of incoherent carriers generated by MIR driving field for HHG**

The calculations in sections VI and VII revealed the significance of EID in the HHG process. Here, we evaluate the impact of incoherent carriers generated by a strong MIR pulse itself on HHG through EID. Figure 8(c) shows the total carrier density calculated without initial carriers and the corresponding evolution of the dephasing time $\hbar/\Gamma$. According to our simulation, an MIR pulse with a peak electric field of 8.3 MV cm$^{-1}$ creates carriers with a density of approximately $7 \times 10^{12}$ cm$^{-2}$ at the center of the pulse duration. This carrier density is comparable to or higher than the density of the doped photo-carriers in the experiment in Fig. 4 (the black dotted line in Fig. 8(c)). These incoherent carriers created by the MIR pulse promote the dephasing process. The calculated dephasing time decreases from 8 fs to 5 fs at the center of the mid-infrared field. This suggests the importance of EID during the MIR pulse irradiation. An MIR driving field with a longer pulse duration may create more incoherent carriers and suppress the high harmonics in the ultraviolet or higher energy region.

The previous sections assumed that EID increases linearly with respect to the carrier density. The validity of this assumption is supported by the experimentally observed MIR intensity dependence, which can be explained as follows. We generally write the carrier density dependent dephasing rate as $\Gamma(N_{\text{MIR}}(I_{\text{MIR}}) + N_{\text{NIR}})$, where $N_{\text{NIR}}$ is the incoherent photo-carrier density and $N_{\text{MIR}}(I_{\text{MIR}})$ is the incoherent carrier density generated by the MIR pulse depending on the intensity $I_{\text{MIR}}$. Since the interband harmonics are exponentially dependent on the dephasing rate as shown in Fig. S3 [18], the ratio between the $n$-th order interband harmonic intensity with ($I^n_{\text{Inter}}(I_{\text{MIR}})$) and without ($I^n_{\text{Inter, 0}}(I_{\text{MIR}})$) photo-carrier doping is expressed as

$$I^n_{\text{Inter}}(I_{\text{MIR}}) = I^n_{\text{Inter, 0}}(I_{\text{MIR}}) \exp\left[-\frac{\Gamma(N_{\text{MIR}}(I_{\text{MIR}}) + N_{\text{NIR}}) - \Gamma(N_{\text{MIR}}(I_{\text{MIR}}))}{\Gamma_n}\right], \quad (17)$$

where $\Gamma_n$ is a normalization parameter for the $n$-th order harmonics. The normalized differential harmonic intensity is given by

$$\frac{\Delta I^n_{\text{Inter}}(I_{\text{MIR}})}{I^n_{\text{Inter, 0}}(I_{\text{MIR}})} = \exp\left[-\frac{\Gamma(N_{\text{MIR}}(I_{\text{MIR}}) + N_{\text{NIR}}) - \Gamma(N_{\text{MIR}}(I_{\text{MIR}}))}{\Gamma_n}\right] - 1. \quad (18)$$

On the other hand, the experiment showed no significant change for the seventh- and higher order harmonics in Fig. 10(d). This can only be explained when the right-hand side of Eq. (18) is independent of the MIR intensity; i.e., the dephasing rate is linearly dependent on the carrier density.

In Fig. 10(c), the fifth-order differential harmonic intensity increases by lowering the MIR intensity. This is because $N_{\text{NIR}}$ becomes relatively larger than $N_{\text{MIR}}(I_{\text{MIR}})$ at lower MIR intensity and results in a relatively large enhancement in the intraband harmonics, which is significant at the fifth order. This is similar in trend to the experimental results in Fig. 10(d) and supports the idea of including the initial carrier distribution in the simulation.

The above considerations point to the necessity of including the photo-carrier doping effects discussed in section VII and suggest the crucial role in HHG of incoherent carriers generated by the MIR pulse itself through EID.

**IX. Screening of excitonic Coulomb interaction at high carrier density regime**

Here, we qualitatively discuss the effect of screening on the excitonic Coulomb interaction at high carrier densities. The calculated initial-carrier-density dependence for the fifth-order harmonics in Fig. 10(a) deviates from the observed dependence in Fig. 10(b) at high photo-carrier densities. The calculated dependence monotonically increases with increasing carrier density, but the experimental results reach a maximum value at a certain density and start to decrease after that. These results should be able to be reproduced by considering the screening effect. The excitonic Coulomb interaction should be gradually screened with increasing photo-carrier density. The harmonic intensity is relatively suppressed at the limit where the Coulomb interaction is completely screened by the electron-hole plasma (see Fig. S4), as a recent work indicated [48]. Thus, our results suggest that the screening of the excitonic-Coulomb interaction overwhelms the effect of the increased intraband current at a photo-carrier density around $N \sim 10^{13}$ cm$^{-2}$, leading to suppression of the high harmonic yields.

The enhancement in below-gap harmonics observed in our study was not observed in the related work [18]. This discrepancy may be explained by differences in the detailed experimental conditions regarding the number of carriers excited by the MIR pulse. The screening of the excitonic Coulomb interaction is dependent on the total carrier density excited by the MIR and NIR pulses. If enough carriers are excited by the MIR pulse so that the screening becomes significant, the harmonics should monotonically decrease as a result of doping additional photo-carriers. One possible reason for the discrepancy is that the longer MIR pulse duration in Ref. [18] results in accumulation of more excited carriers than ours.

The above discussion suggests the importance of many-body interactions in HHG, which should be addressed in future investigations.

## X. Conclusion

We studied the effects of incoherent electron-hole pairs on high harmonic generation in an ideally thin semiconductor under a strong MIR field. We found that, due to photo-carrier doping, the harmonic intensity changes positively and negatively below and above the absorption edge. To describe the photo-carrier doping effect, we performed numerical simulations considering the electron-electron scattering process, the excitonic Coulomb interaction, and the photo-carrier distributions. They showed that photo-carriers enhance the intraband current relevant to lower-order harmonics, which is relatively suppressed by momentum relaxation. We also showed that photo-carriers significantly suppress the interband current that contributes to higher-order harmonics above the absorption edge. We clarified that the main suppressive effect is not the phase-space filling but rather the EID. Our work revealed that many-body effects, such as electron-electron scattering, play a crucial role in the solid-state HHG. These findings provide us with a deeper understanding of high harmonic spectroscopy in solids and on the generation of broadband light from solids.

## Acknowledgements

We thank Y. Shinohara for fruitful discussions. This work was supported by Grants-in-Aid for Scientific Research (S) (Grant Nos. JP17H06124 and JP21H05017) and a JST ACCEL Grant (No. JPMJMI17F2). K.N. was supported by a JSPS fellowship (Grant No. JP20J14428). K.U. is thankful for a Grant-in-Aid for Young Scientists (Grant No. 19K14632). S.K. was supported by MEXT Quantum Leap Flagship Program (MEXT Q-LEAP) (Grant No. JPMXS0118067634). Y.M. acknowledges support from a Grant-in-Aid for Transformative Research Areas (Grant No. JP21H05232, JP21H05234). We thank H. Shimizu for his support on sample fabrication.

# Supplementary Material
# Effect of incoherent electron-hole pairs on high harmonic generation in an atomically thin semiconductor

Kohei Nagai[1], Kento Uchida[1], Satoshi Kusaba[1], Takahiko Endo[2], Yasumitsu Miyata[2], and Koichiro Tanaka[1,3]

[1]Department of Physics, Graduate School of Science, Kyoto University, Sakyo-ku, Kyoto 606-8502, Japan.
[2]Department of Physics, Tokyo Metropolitan University, Hachioji, Tokyo 192-0397, Japan.
[3]Institute for Integrated Cell-Material Sciences, Kyoto University, Sakyo-ku, Kyoto, 606-8501, Japan

**SI. Derivation of the fitting function for degenerate pump and probe experiment**

We derive the equation for the fitting of exciton-exciton annihilation (EEA) and absorption saturation. First we consider a rate equation for two-level systems. The rate equation used for the modeling is

$$\frac{dN_1}{dt} = -k_A N_1^2 - \alpha \frac{I_p(t)}{\hbar \omega_N}(N_1 - N_2), \tag{S1}$$

where $k_A$ is the EEA rate, $N_1$ and $N_2$ are the population per unit area in level 1 and 2, respectively, $I_p(t)$ is the intensity of the near-infrared (NIR) pulse, and $\alpha$ is the optical cross section of the NIR photon on the monolayer. The population $N_1$ corresponds to the one of the exciton. The photon energy of the NIR pulse is assumed to be resonant with the two-level systems.

First, we calculate the absorbance of the NIR probe beam with the initial population $N_1 = N_i$. The absorbance is defined by the change between the population before and after the probe pulse irradiation. Since the EEA is typically slower than the NIR pulse width ($\sim 100$ fs), we will neglect the first term in the Eq. (S1) and solve

$$\frac{dN_1}{dt} = -\alpha \frac{I_p(t)}{\hbar \omega_N}(N_1 - N_2) \tag{S2}$$

to calculate $N_1$ just after the probe pulse irradiation. We assume $N_1 + N_2 = N_T$, where the $N_T$ is the total number of the two-level system. By defining $\Delta N(t) = N_2 - N_1$, we get

$$\Delta N(t) = \Delta N(0) \exp\left(-\frac{F_p}{F_S}\right), \tag{S3}$$

where

$$F_p = \int_{-\infty}^{t} dt\, I_p(t) \tag{S4}$$

is the NIR probe fluence and $F_S = \hbar\omega_N/2\alpha$ is the saturation fluence. The absorbance (optical density) is calculated by

$$A(N_i) = \left.\frac{dN_1 \hbar\omega_N}{dF_p}\right|_{F_p=0} = -\frac{1}{2}\left.\frac{d\Delta N \hbar\omega_N}{dF_p}\right|_{F_p=0} = \frac{\Delta N \hbar\omega_N}{2F_S}. \tag{S5}$$

Here, $A = A(0) = N_T \hbar\omega_N / 2F_S$ corresponds to the absorbance without the pump pulse.

On the other hand, in the case of the transparent substrate, the difference between the transmission signals for the monolayer on the substrate and for bare substrate is proportional to the absorbance of the monolayer [49],

$$\frac{T - T_s}{T_s} = -\frac{2}{n_S + 1} A. \tag{S6}$$

Here, $T(T_s)$ is the transmittance for the monolayer WSe$_2$ on the substrate (bare substrate) and $n_S = 1.76$ is the refractive index of the sapphire substrate at 1.63eV [32]. When the absorbance of the monolayer is small, the differential transmittance due to the photo-carrier doping is given by

$$\frac{\Delta T}{T} \approx -\frac{2}{n_S + 1}\Delta A = -\frac{2}{n_S + 1}\big(A(N_i) - A(0)\big) = \frac{\hbar\omega_N}{F_S}\frac{2N_1}{n_S + 1}. \tag{S7}$$

Next we calculate the carrier density excited by the pump NIR pulse. It can be calculated by replacing the fluence of the probe pulse in Eq. (S3) with that of the pump pulse $F_N$. By assuming $N_i = 0$, we obtain the population before the relaxation process:

$$N_1(0) = \frac{AF_S}{\hbar\omega_N}\left(1 - \exp\left(-\frac{F_N}{F_S}\right)\right). \tag{S8}$$

The population at the time delay $t$ is calculated by considering EEA process:

$$\frac{dN_1}{dt} = -k_A N_1^2. \tag{S9}$$

By solving this equation, we obtain

$$N_1(t) = \frac{N_1(0)}{1 + k_A N_1(0) t}. \tag{S10}$$

Finally, we obtain the differential transmittance at the time delay $t$ from Eqs. (S7), (S8), (S10) as follows:

$$\frac{\Delta T}{T} = \frac{\hbar\omega_N}{F_S}\frac{2}{n_S+1}\frac{N_1(0)}{1+k_A N_1(0)t} = \frac{2A}{n_S+1}\frac{1-\exp\left(-\frac{F_N}{F_S}\right)}{1+k_A\frac{AF_S}{\hbar\omega_N}\left(1-\exp\left(-\frac{F_N}{F_S}\right)\right)t}. \tag{S11}$$

The global fitting in Fig. 3 in the main paper was performed by using Eq. (S11) with common free parameters $A, k_A, F_S$. The fitting results are summarized in the Table S1.

Table S1 Fitting results for Figure 3 in main paper

| parameter | fitting result |
|---|---|
| $A$ | $0.03220 \pm 0.00070$ |
| $k_A$ | $(2.46 \pm 0.10) \times 10^{-2}$ cm$^2$ s$^{-1}$ |
| $F_S$ | $(1.014 \pm 0.039) \times 10^2$ μJ cm$^{-2}$ |

Table S2 Fitting results for the excitonic wave function with Eq. (12) in main paper

| parameter | fitting result |
|---|---|
| $c_0$ | 0.00031607156 |
| $c_1$ | 0.36395591 |
| $c_2$ | 1.9432212 |
| $c_3$ | 5.5144512e-5 |
| $\sigma$ | 2.0528135 |

**SII. Transformation of semiconductor Bloch equations**

The semiconductor Bloch equations (SBEs) written in the main text are transformed from the SBEs known as partial differential equations [5,7,27,28,34]. We start with the following form:

$$\frac{\partial}{\partial t} P_\mathbf{k} = -\frac{i}{\hbar}\left[e_g(\mathbf{k}) + ie\mathbf{E}(t)\cdot\frac{\partial}{\partial \mathbf{k}} - i\Gamma\right]P_\mathbf{k} - i\omega_{R,\mathbf{k}}(2n_\mathbf{k} - 1), \quad (S12)$$

$$\frac{\partial}{\partial t} n_\mathbf{k} = -2\mathrm{Im}[\omega_{R,\mathbf{k}} P_\mathbf{k}^*] + \frac{e\mathbf{E}(t)}{\hbar}\cdot\frac{\partial}{\partial \mathbf{k}} n_\mathbf{k} - \frac{1}{\hbar}\gamma_m[n_\mathbf{k} - n_{\mathbf{k},eq}], \quad (S13)$$

where $P_\mathbf{k}$ and $n_\mathbf{k}$ are the interband coherence and the population at the crystal momentum $\mathbf{k}$, respectively. The renormalized single-particle bandgap energy $e_g(\mathbf{k})$ and the generalized Rabi frequency $\omega_{R,k}$ are written as

$$e_g(\mathbf{k}) = \varepsilon_g(\mathbf{k}) - 2\sum_{\mathbf{q}\neq\mathbf{k}} V_{\mathbf{k}-\mathbf{q}}\, n_\mathbf{q}, \quad (S14)$$

$$\omega_{R,\mathbf{k}} = \frac{1}{\hbar}\left[\mathbf{d}_{cv}\cdot\mathbf{E}(t) + \sum_{\mathbf{q}\neq\mathbf{k}} V_{\mathbf{k}-\mathbf{q}}\, P_\mathbf{q}\right], \quad (S15)$$

respectively [34]. Here, a symmetric population in the conduction and valence bands $n_\mathbf{K} = n_{c,\mathbf{K}} = 1 - n_{v,\mathbf{K}}$ is assumed. Dephasing and momentum relaxation are introduced phenomenologically with parameters $\Gamma$ and $\gamma_m$.

To transform Eqs. (S12)-(S15) into ordinary differential equations, we change the variables from ($\mathbf{k}$, $t$) to new independent variables ($\mathbf{K}$, $t'$) as

$$\begin{pmatrix}\mathbf{k}\\t\end{pmatrix} \mapsto \begin{pmatrix}\mathbf{K}\\t'\end{pmatrix} = \begin{pmatrix}\mathbf{k} - e\mathbf{A}(t)/\hbar\\t\end{pmatrix}.$$

In this coordinate transformation, the differential operators are replaced by

$$\begin{pmatrix}\frac{\partial}{\partial \mathbf{k}}\\\frac{\partial}{\partial t}\end{pmatrix} = \begin{pmatrix}\frac{\partial}{\partial \mathbf{K}}\\\frac{e}{\hbar}\mathbf{E}(t') \cdot \frac{\partial}{\partial \mathbf{K}} + \frac{\partial}{\partial t'}\end{pmatrix}.$$

The SBEs in the coordinates of ($\mathbf{K}$, $t'$) take the following forms:

$$\frac{\partial}{\partial t'}\tilde{P}_\mathbf{K} = -\frac{i}{\hbar}\left[e_g\left(\mathbf{K} + \frac{e}{\hbar}\mathbf{A}(t')\right) - i\Gamma\right]\tilde{P}_\mathbf{K} - i\omega_{R,\mathbf{K}+\frac{e}{\hbar}\mathbf{A}(t')}(2\tilde{n}_\mathbf{K} - 1), \tag{S16}$$

$$\frac{\partial}{\partial t'}\tilde{n}_\mathbf{K} = -2\mathrm{Im}\left[\omega_{R,\mathbf{K}+\frac{e}{\hbar}\mathbf{A}(t')}\tilde{P}_\mathbf{K}^*\right] - \frac{1}{\hbar}\gamma_m\left[\tilde{n}_\mathbf{K} - n_{\mathbf{K}+\frac{e}{\hbar}\mathbf{A}(t'),eq}\right], \tag{S17}$$

$$e_g\left(\mathbf{K} + \frac{e}{\hbar}\mathbf{A}(t')\right) = \varepsilon_g\left(\mathbf{K} + \frac{e}{\hbar}\mathbf{A}(t')\right) - 2\sum_{\mathbf{Q}\neq\mathbf{K}}V_{\mathbf{K}-\mathbf{Q}}\tilde{n}_\mathbf{Q}, \tag{S18}$$

$$\omega_{R,\mathbf{K}+\frac{e}{\hbar}\mathbf{A}(t')} = \frac{1}{\hbar}\left[\mathbf{d}_{cv} \cdot \mathbf{E}(t') + \sum_{\mathbf{Q}\neq\mathbf{K}}V_{\mathbf{K}-\mathbf{Q}}\tilde{P}_\mathbf{Q}\right], \tag{S19}$$

with $\tilde{P}_\mathbf{K} = P_\mathbf{k}$ and $\tilde{n}_\mathbf{K} = n_\mathbf{k}$.

**SIII. Time-evolution of intraband harmonics in mid-infrared pulse duration**

Figure S1 shows the time-windowed Fourier spectrum of the intraband harmonics. The third (0.78 eV) and fifth (1.3 eV) order harmonics have intensity maxima after the center of the mid-infrared pulse. This indicates that real carriers are mostly created at the center and the oscillation of the real carriers in the anharmonic band structure produces the lower order intraband harmonics. On the other hand, higher order harmonics have their intensity maxima at the center. This indicates that the higher order intraband harmonics come from the temporally oscillating population following the interband polarization [15]. In our study, this contribution is smaller than the interband harmonics as shown in Fig. 9 (a).

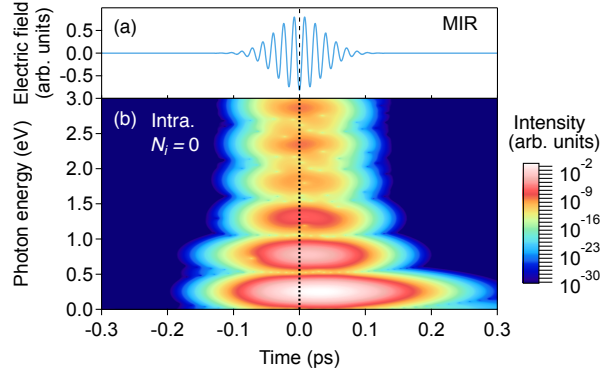

Fig. S1 Time-windowed Fourier spectrum of intraband high harmonic generation. (a) Computed mid-infrared electric field. (b) Time-windowed Fourier spectrum without initial carriers ($N_i = 0$). Temporal Gaussian window of two cycles of mid-infrared field (full-width at half maximum) is used.

**SIV. Effect of the momentum relaxation on intraband harmonics and dephasing on interband harmonics**

Figure S2 shows the results of the numerical calculation of the intraband harmonics without (red) and with (blue) momentum relaxation of $\hbar/\gamma_m = 19$ fs with the initial carrier density $N_i = 3.7 \times 10^{12}$ cm$^{-2}$. In this calculation, the dipole moment $d_{cv}$ is set to 0 to neglect the effect of interband transitions for simplicity. The spectra show exponential decays with increasing the harmonic order, which is derived from the Bloch oscillation of the initial carriers [15]. The dependence of the each order intraband harmonic on the momentum relaxation rate is shown in Fig. S2(b). The momentum relaxation rate up to a moderate rate exponentially suppresses the intraband harmonics. Figure S3 represents the dephasing rate dependence of the interband harmonics. The interband harmonics exponentially decrease with respect to the dephasing rate.

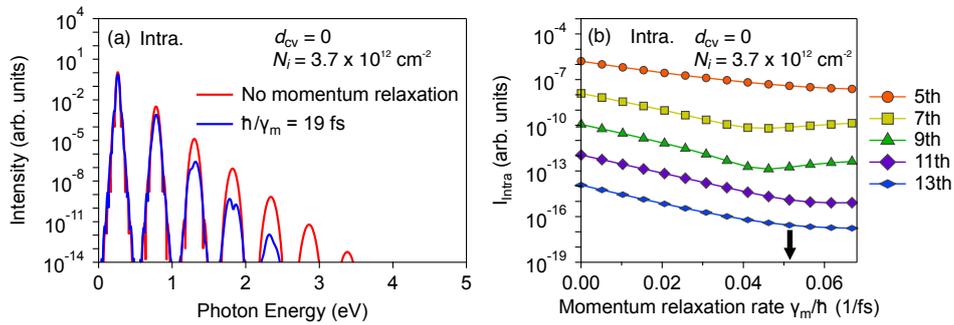

Fig. S2 (a) Intraband harmonic spectra calculated with an initial carrier density of $N_i = 3.7 \times 10^{12}$ cm$^{-2}$ and $d_{cv} = 0$. Two spectra represent the results with $\hbar/\gamma_m = \infty$ (red) and $\hbar/\gamma_m = 19$ fs (blue). (b) Momentum-relaxation-rate dependence of intraband harmonic intensity. The black arrow indicates the rate of 19 fs, which is used for the full calculation.

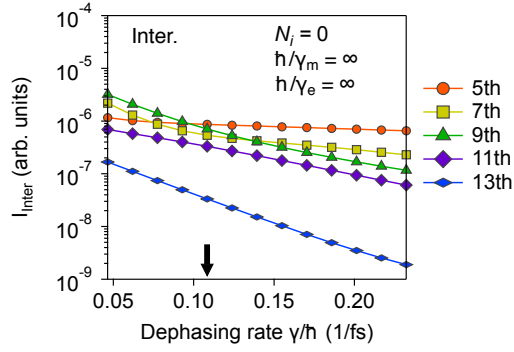

Fig. S3. The dependence of interband harmonic intensity on dephasing rate without initial carriers ($N_i = 0$), and $\hbar/\gamma_m = \infty$, $\hbar/\gamma_e = \infty$. The black arrow represents the dephasing rate corresponding to $\hbar/\gamma = 8$ fs, which is used for all the other calculations.

**SV. Impact of excitonic Coulomb interaction on high harmonic intensity**

Figure S4 shows the high harmonic spectra calculated with and without the excitonic Coulomb interaction. The excitonic Coulomb interaction enhances the harmonic intensity by several orders of magnitude. This is caused by the enhanced interband transition through the large excitonic absorption at the absorption edge. The calculation without the excitonic Coulomb interaction ($V_q = 0$) corresponds to the limit where the excitonic Coulomb interaction is completely screened by the electron-hole plasma. Thus, the screening of the excitonic Coulomb interaction decreases the harmonic intensity.

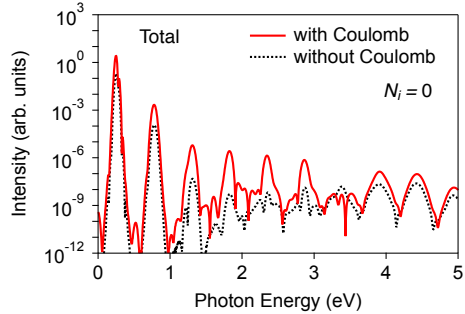

Fig. S4. High harmonic generation spectra calculated with (red solid line) and without (dotted line) excitonic Coulomb interaction. The harmonic spectra are calculated without initial carriers ($N_i = 0$) and with the parameters $\hbar/\gamma = 8$ fs, $\hbar/\gamma_m = 19$ fs, and $\hbar/\gamma_e = 7.9 \times 10^{13}$ fs cm$^{-2}$. The spectrum without Coulomb interaction is calculated by setting $V_q = 0$.

**SVI. Evaluation of the sample quality**

To evaluate the sample degradation under the strong MIR irradiation, we measured the PL spectra at 293K and 6K before and after the HHG measurements as shown in Figs. S5(a) and S5(b). At both temperatures, we observed a clear decrease in the PL intensity due to the MIR irradiation. Before the MIR irradiation (red), PL peaks, including the exciton ($X_0$), charged exciton ($X^*$), and possibly the intervalley exciton or dark exciton phonon replica peaks ($X_i$), are obtained at 6K [50]. After completion of all experimental procedures (blue), only the broad PL peak of the localized excitons (Loc.) is observed at 6K and the PL intensity decreases drastically at 293 K. These results indicate that defect formation should take place extensively by the strong MIR pulses. However, it is noteworthy that we found no clear decrease in HHG efficiency throughout our HHG measurement. This is because the efficiency of PL, which is accompanied by the diffusion process, is more sensitive to the defect density than the efficiency of HHG.

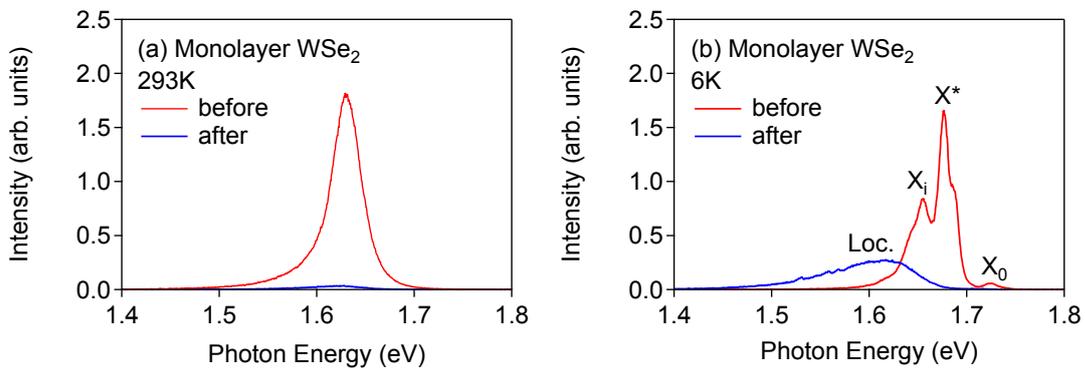

Fig. S5 Photoluminescence (PL) spectra of monolayer WSe$_2$ before (blue) and after (red) the irradiation with strong mid-infrared (MIR) pulses at (a) 293K and (b) 6K. The sample was loaded in a flow liquid helium cryostat and the spectra were obtained with a commercial micro-PL spectrometer (NanoFinder30, Tokyo Instruments Inc.). The PL intensity decreases from before to after the MIR irradiation at both temperatures. The PL spectrum at 6K before the MIR irradiation exhibits sharper peaks than that after the MIR irradiation, including free exciton ($X_0$), charged exciton ($X^*$), and possibly intervalley exciton or dark exciton phonon replica ($X_i$). The PL spectrum after the MIR irradiation is dominated by the broad peak of the localized excitons (Loc.).